\documentclass[11pt,leqno]{article}

\usepackage{graphicx} 

\usepackage[dvipsnames]{color} 
\definecolor{Red}{named}{Red}

\usepackage{amsmath} 
\usepackage{latexsym,epsfig,bm,amssymb} 
\usepackage{color} 
\usepackage{amsthm,mathrsfs}

 \usepackage{mathptmx}

\newcommand{\nc}{\newcommand} 
\nc{\dps}{\displaystyle} 

\newtheorem{theorem}{Theorem}

\newcommand{\loota}{\hbox{\enspace{\vrule height 7pt depth 0pt width 
      7pt}}} 
\nc{\RR}{\mbox{\rm I$\!$R}} 
 
 \newcommand{\beqn}{\begin{eqnarray}} 
 \newcommand{\eeqn}{\end{eqnarray}} 
 \newcommand{\be}{\begin{equation}} 
 \newcommand{\ee}{\end{equation}} 
 \newcommand{\ba}{\begin{array}} 
 \newcommand{\ea}{\end{array}} 
 \newcommand{\pa}{\partial} 
 \newcommand{\re}{\ref} 
 \newcommand{\ci}{\cite} 
 \newcommand{\ds}{\displaystyle} 
 \newcommand{\la}{\label} 
 \newcommand{\bfr}{\begin{flushright}} 
 \newcommand{\efr}{\end{flushright}} 
  
 \newcommand{\rIm}{{\rm Im\5}} 
  
\newcommand{\bfl}{\begin{flushleft}} 
\newcommand{\efl}{\end{flushleft}} 
\newcommand{\fr}{\frac}

\newcommand{\ov}{\overline} 
\newcommand{\st}{\stackrel}

\newcommand{\toLp}{\st{L^p}\longrightarrow} 
 
\newcommand{\toLd}{\st{L^2}\longrightarrow}

\def\longrightharpoonup{\relbar\joinrel\rightharpoonup} 
\newcommand{\tow}{\st{L^2_w}{\longrightharpoonup}}

\newcommand{\dist}{{\rm dist}} 

\newcommand{\bo}{{\hfill\loota}} 
 
\renewcommand{\Pr}{{\bf Proof.~}}

\newcommand{\bba}{{\bf a}} 
 
\newcommand{\n}{{\bf n}} 
\newcommand{\bb}{{\bf b}}

\newcommand{\x}{{\bf x}} 
\newcommand{\y}{{\bf y}}

\newcommand{\bk}{{\bf k}} 
 
\newcommand{\cm}{{\rm m}}

 \newcommand{\cM}{{\mathcal M}}

\newcommand{\E}{{\cal E}}

\newcommand{\ccT}{{\cal T}}

\newcommand{\cX}{{\mathcal X}}

\newcommand{\ve}{\varepsilon}

\newcommand{\De}{\Delta} 
\newcommand{\de}{\delta} 
 
\newcommand{\al}{\alpha} 
 
\newcommand{\Ga}{\Gamma} 
\newcommand{\si}{\sigma} 
\newcommand{\om}{\omega} 
 
\newcommand{\na}{\nabla} 
 
\newcommand{\lam}{\lambda} 
\newcommand{\Lam}{\Lambda} 
 \newcommand{\h}{{\hbar}}

\newcommand{\5}{{\hspace{0.5mm}}} 
\newcommand{\C}{{\mathbb C}} 
\newcommand\R{{\mathbb R}} 
\newcommand\Z{{\mathbb Z}}

\newtheorem{qtheorem}{QTheorem}[section]

\newtheorem{defin}[theorem]{Definition} 
 
\newtheorem{lemma}[theorem]{Lemma} 
\newtheorem{example}[theorem]{Example} 
\newtheorem{exercice}[theorem]{Exercise} 
\newtheorem{remark}[theorem]{Remark} 
\newtheorem{remarks}[theorem]{Remarks} 
\newtheorem{cor}[theorem]{Corollary} 
\newtheorem{pro}[theorem]{Proposition} 
 
\newtheorem{coms}[theorem]{Comments}

\newcommand{\bd}{\begin{defin}} 
 \newcommand{\ed}{\end{defin}} 
\newcommand{\bt}{\begin{theorem}} 
 \newcommand{\et}{\end{theorem}} 
\newcommand{\bqt}{\begin{qtheorem}} 
 \newcommand{\eqt}{\end{qtheorem}}

\newcommand{\bp}{\begin{pro}} 
 \newcommand{\ep}{\end{pro}} 
 
\newcommand{\bl}{\begin{lemma}} 
 \newcommand{\el}{\end{lemma}} 
\newcommand{\bc}{\begin{cor}} 
 \newcommand{\ec}{\end{cor}} 
 
\newcommand{\bex}{\begin{example}} 
 \newcommand{\eex}{\end{example}} 
\newcommand{\bexs}{\begin{examples}} 
 \newcommand{\eexs}{\end{examples}}

\newcommand{\bexe}{\begin{exercice}} 
 \newcommand{\eexe}{\end{exercice}}

\newcommand{\br}{\begin{remark} } 
 \newcommand{\er}{\end{remark}} 
\newcommand{\brs}{\begin{remarks}} 
 \newcommand{\ers}{\end{remarks}}

\newcommand{\bcoms}{\begin{coms}} 
\newcommand{\ecoms}{\end{coms}}

\begin{document}

% \hspace{50mm} arXiv: 1310.3084v1 [math-ph] 11 Oct 2013

~ \vspace{20mm} 
\begin{center} 
%\hspace{10cm} 
%{\huge THIS IS DRAFT} 
%\\~ 
%\\ 
{\huge\bf  On  crystal ground state in the
\bigskip\\ 
Schr\"odinger--Poisson model with point ions} 
\\~ 
\\~ 
\\ 
\vspace{15mm} 
{\large A.\,I.~Komech 
\footnote{The research was carried out at the IITP RAS at the expense of the Russian
Foundation for Sciences (project  14-50-00150).
}}\\ 
{\it Faculty of Mathematics of  Vienna  University and\\ 
Institute for Information Transmission Problems RAS\\}
alexander.komech@univie.ac.at\\

\bigskip\bigskip\bigskip\bigskip\bigskip

\end{center} 
 
\begin{abstract} 
A~space-periodic ground state is shown to exist for lattices of 
point
ions in $\R^3$ coupled to the Schr\"odinger and scalar fields. 
The coupling requires the renormalization due to the 
singularity of the Coulomb selfaction.
The ground state is constructed by minimization of the renormalized 
energy per cell. This  energy is bounded from below 
when the charge of each ion is positive.
The  elementary cell is necessarily neutral.

\end{abstract}


%%\tableofcontents 

%%%%%%%%%%%%%%%%%%%%%%%%%%%%%%%%%%%%%%%%%%%%%%%% 
%%%%%%%%%%%%%%%%%%%%%%%%%%%%%%%%%%%%%%%%%%%%%% 
 
 \newpage 
\setcounter{equation}{0} 
\setcounter{section}{-1} 
\setcounter{section}{0} 
\section 
{Introduction} 
We consider  $3$-dimensional crystal lattices in $\R^3$, 
\be\la{Ga3} 
\Ga:=\{\x(\n)=\bba_1 n_1+
\bba_2 n_2+\bba_3 n_3: \n=(x_1,x_2,x_3)\in\Z^3  \}, 
\ee 
 $\bba_k\in\R^3$ are linearly independent periods. 
Born and Oppenheimer \ci{BO} developed 
the  quantum dynamical approach to the crystal structure, 
separating the motion of  `light electrons' and of `heavy ions'. 
As an extreme form of this separation, 
the ions could be 
considered as classical nonrelativistic particles governed by the Lorentz equations 
neglecting the magnetic field, 
while the electrons could be  described by the Schr\"odinger 
equation neglecting the electron spin. 
The scalar potential 
is the solution to the corresponding Poisson equation. 

We consider the crystal with $N$ ions per cell.
Let  $\si_j(\y)=|e|Z_j\de(\y)$ be
the charge density 
and $M_j>0$ the mass
of the corresponding ion, $j=1,...,N$.
Then the coupled equations 
 read 
\beqn 
i\h\dot\psi(\x,t)&=&-\fr{\h^2}{2\cm}\De\psi(\x,t)+e\phi(\x,t)\psi(\x,t), ~~~~~\x\in\R^3,
\la{LPS1} 
\\ 
\nonumber\\ 
\qquad \Bigl [\ds\fr1{c^2}\pa_t^2-\De\Bigr] 
\phi(\x,t)&=&\rho(\x,t):= 
\sum_{j=1}^N\5\sum _{\n\in\Z^3} 
\si_j(\x-\x(\n)-\x_j(\n,t))+e|\psi(\x,t)|^2, ~~~~~\x\in\R^3,\la{LPS2} 
\\ 
\nonumber\\ 
M_j\ddot\x_j(\n,t) 
&=&-|e|Z_j\na\phi_{\n,j}(\x(\n)+\x_j(\n,t)),
\quad \n\in\Z^3,\quad 
j=1,\dots,N. 
\la{LPS} 
\eeqn 
Here 
$e<0$ is the electron charge, 
$\cm$ is its mass, 
$\psi(\x,t)$ denotes the 
wave function of the electron field, and 
$\phi(\x,t)$ is the potential of the 
scalar field generated by the ions and the electrons. 
Further, 
$(\cdot,\cdot )$ 
stands for the Hermitian scalar product in the Hilbert 
space $L^2(\R^3)$, and
\be\la{phinj}
\na\phi_{\n,j}(\x(\n)+\x_j(\n,t)):=\na_\y\left[\phi(\x(\n)+\x_j(\n,t)+\y)-\fr{|e|Z_j}{4\pi |\y|}\right]\Bigg|_{\y=0}
\ee
All derivatives here and below 
are understood in the sense of distributions. 
The system is  nonlinear and translation invariant, 
i.e., $\psi(\x-\bba,t)$, $\phi(\x-\bba,t)$, 
$\x_j(\n,t)+\bba$ is also a~solution for any $\bba\in\R^3$ . 

 %%%%%%%%%%%%%%%%%%%%%
 A dynamical quantum 
 description of the solid state as  many-body system
is not rigorously established yet
 (see Introduction of \ci{LAK} and Preface of \ci{Peierls}).
 %%%%%%%%%%%%%%%%%%%%%
Up to date rigorous results concern only the ground state in different models
(see below).

The classical "one-electron" theory of Bethe-Sommerfeld,  
based on  periodic Schr\"odinger equation, does not take into account
oscillations of  ions.
Moreover, the choice of the periodic potential in this theory is very problematic, and 
corresponds to a fixation of the ion positions which are unknown.
 
The system (\re{LPS1})--(\re{LPS})  eliminates these  difficulties
though  it does  not respect the electron spin like the periodic Schr\"odinger equation.
To remedy this deficiency we should replace the Schr\"odinger equation by 
the Hartree--Fock equations as the next step to more realistic model.
However, we expect that the  techniques developed for the system
(\re{LPS1})--(\re{LPS})
will be useful also for more realistic  dynamical models of crystals.
These goals were our main motivation in writing this paper.
%%%%%%%%%%%%

 Here, we make the first step  proving the existence
 of the ground state, which is
 a~$\Ga$-periodic stationary solution 
$\psi^0(\x)e^{-i\om ^0t}$,
$\phi^0(\x)$, 
$\ov\x=(\x^0_1,~\dots,\x^0_N)$  to the system 
(\re{LPS1})--(\re{LPS}): 
\beqn\la{LPS3} 
\h\om^0\psi^0(\x)&=&-\fr{\h^2}{2\cm}\De\psi^0(\x)+e\phi^0(\x)\psi^0(\x), 
~~~~\x\in T^3,
\\ 
\nonumber\\ 
-\De\phi^0(\x)&=&\rho^0(\x):= 
\si^0(\x)+e|\psi^0(\x)|^2, ~~~~~~\x\in T^3, 
\la{LPS4} 
\\ 
\nonumber\\ 
\la{LPS3g} 
0&=&-|e|Z_j \na\phi_{\n,j}^0(\x^0_j), \qquad  j=1,\dots,N. 
\eeqn 
Here, $T^3:=\R^3/\Ga$ denotes  the `elementary  cell' of the crystal, 
$\langle\cdot,\cdot\rangle$ 
stands for the Hermitian scalar product in the complex Hilbert space $L^2(T^3)$ 
and its different extensions, and 
\be\la{rrr} 
\si^0(\x):= 
\sum_{j=1}^N \si_j(\x-\x^0_j),\qquad
\si_j(\y):=|e|Z_j\de(\y).
\ee  
The right hand side  of (\re{LPS3g}) is defined similarly to (\re{phinj}):
\be\la{phinj0}
\na\phi_{\n,j}^0(\x_j^0):=\na_\y\left[\phi(\x_j^0+\y)-\fr{|e|Z_j}{4\pi |\y|}\right]\Bigg|_{\y=0}
\ee
The system (\re{LPS3})--(\re{LPS3g}) is translation invariant similarly to (\re{LPS1})--(\re{LPS}).
Let us note that 
$\om^0$ should be real since
$\rIm\om^0\ne 0$ means an instability of the ground state:
the decay 
as $t\to\infty$
in the case $\rIm\om^0 < 0$ and the explosion  if $\rIm\om^0 > 0$.
We have
\be\la{intro}
\int_{T^3}\si^0(\x)d\x=Z|e|,\qquad Z:=\sum_j Z_j. 
\ee 
The total charge per cell should be zero (cf. \ci{BBL2003}): 
\be\la{neu10} 
\int_{T^3} \rho^0(\x)d\x= 
\int_{T^3} [\si^0(\x) +e|\psi^0(\x)|^2]d\x=0. 
\ee 
This neutrality condition
follows directly from equation (\re{LPS4}) by integration using 
$\Ga$-periodicity of  $\phi^0(\x)$.
Equivalently,
the neutrality condition can be written as the normalization
\be\la{neuZ} 
\int_{T^3} |\psi^0(\x)|^2d\x 
=Z. 
\ee 
Our main condition is the following:
\be\la{Zp}
 \mbox{\bf Positivity condition:} \qquad\quad Z_j>0,\quad j=1,..., N.\qquad\qquad\qquad\qquad
\ee
Let us comment on our approach. 
The neutrality condition (\re{neuZ}) 
defines the submanifold $\mathcal M$ in the space 
$H^1(T^3)\times (T^3)^N$
of 
space-periodic configurations $(\psi^0,\ov\x^0)$. 
We construct a~ground state as a~minimizer 
over $\mathcal M$
of the energy per cell 
(\re{HamsT}). 
Previously we have established similar results  \ci{K-2014} for the crystals 
with 1D, 2D and 3D lattices
of smeared ions in $\R^3$. 
Our main novelties in the present paper are the following.
\medskip \\
I. We extend our results \ci{K-2014} to the point ions
subtracting the infinite selfaction in the 
renormalized equations.
\medskip \\
II.  We renormalize the energy per cell subtracting the infinite Coulomb selfaction of the point
ions. 
\medskip \\
III. We prove the bound from below for the renormalized energy 
under the novel assumption (\re{Zp}) of the positivity for the charge of each ion.
 \medskip \\
The minimization strategy ensures the existence of a~ground state for any
 lattice~\eqref{Ga3}. One could expect
 that a stable lattice should provide a local minimum of
 the energy per  cell
 for fixed $N$ and  $Z_j$, but this is still an open
problem.
\medskip 
 
Let us comment on related works. 
For atomic systems in $\R^3$, a~ground state was constructed by Lieb, Simon  and P.~Lions 
in the case of the Hartree and Hartree--\allowbreak Fock models
\ci{LS1977,Lions1981, Lions1987}, and 
by Nier
for the Schr\"odinger--\allowbreak Poisson  model \ci{Nier93}. 
The Hartree--\allowbreak Fock dynamics 
for molecular  systems in $\R^3$
has been constructed
by Canc\`es and Le Bris \ci{CB}.

A mathematical theory of the stability of  matter 
started from the pioneering works of 
Dyson, Lebowitz, Lenard, Lieb and others 
 for the Schr\"odinger many body model 
\ci{Dyson1967, Lieb2005, LL1972, Lieb2009}; 
see the survey in~\ci{Lemm}. 
Recently, the theory was extended to the 
quantized Maxwell field \ci{LL2005}.

These results and methods were developed 
last two decades
by Blanc, Le Bris, Catto, P. Lions and others 
to justify 
the  thermodynamic limit for the Thomas--Fermi and Hartree--\allowbreak Fock 
models 
with space-periodic  ion arrangement 
\ci{BBL2007,CBL1996,CBL1998,CBL2001} 
and to construct the corresponding space-periodic ground states 
\ci{CBL2002}, 
see the survey and further references in \ci{BL2005}. 
 
Recently, Giuliani, Lebowitz and Lieb have established 
the periodicity of the thermodynamic limit 
in 1D local mean field model 
without the assumption of periodicity of the ion arrangement~\ci{GLL2007}. 
 
Canc\`es and others studied  short-range perturbations 
of the  Hartree--\allowbreak Fock 
model and proved that 
the 
density matrices of the perturbed and  unperturbed 
ground states 
differ by a~compact operator, 
\ci{CL2010,CS2012}. 
 
 %%%%%%%%%%%%%%%%%%%%%%%%%%%%%%%%%%%%%%%
The Hartree--Fock dynamics for infinite particle systems were considered recently 
by Cances and Stoltz \ci{CS2012}, and Lewin and Sabin \ci{LS2014-1}.
In \ci{CS2012}, the well-posedness  is established for 
local perturbations of the periodic ground state density matrix
in an infinite crystal.
However, the  space-periodic nuclear potential 
in the equation \ci[(3)]{CS2012}
is fixed that corresponds to 
the fixed nuclei positions. Thus the back reaction of the electrons onto 
the nuclei is neglected.
In \ci{LS2014-1},  the well-posedness is established for the 
von Neumann equation  with density matrices of infinite trace 
for pair-wise interaction potentials $w\in L^1(\R^3)$. Moreover, the authors  
prove the asymptotic stability of the ground state in 2D case \ci{LS2014-2}.
Nevertheless, the case of  the Coulomb potential for infinite particle systems remains open
since the corresponding generator is infinite.

\medskip 
 
The plan of our paper is as follows. 
In Section 2, we renormalize the energy per cell
and prove that the renormalized energy is bounded from below.
In Section 3, we prove the compactness of the minimizing sequence,
and in Section 4 calculate the energy variation.
In the final Section 5, we prove the Schr\"odinger equation.

\medskip

{\bf Acknowledgments.} 
The author thanks H. Spohn for useful remarks and E. Kopylova for helpful discussions.

\setcounter{subsection}{0} 
\setcounter{theorem}{0} 
\setcounter{equation}{0}

\section{The renormalized energy per cell}

We consider the system  \eqref{LPS3}, (\re{LPS4}) 
for the corresponding functions on the torus  $T^3=\R^3/\Ga$ 
and for $\x_{0j}\5{\rm mod}\5\Ga\in T^3$. 
For $s\in\R$, we denote by $H^s$ the Sobolev 
space on the torus $T^3$, and for $1\le p\le \infty$, we denote by $L^p$ 
the Lebesgue space of  functions on $T^3$.

The ground state will be constructed by minimizing the 
energy  in the cell $T^3$. 
To this aim, we will minimize the energy with respect to 
$\ov\x:=(\x_1,\dots,\x_N)\in (T^3)^N$ and 
$\psi\in H^1$ satisfying the neutrality condition 
(\re{neu10}): 
\be\la{neu1} 
\int_{T^3} \rho(\x)d\x=0,~~~~~~~~~~\rho(\x):=\si(\x) +\nu(\x),
\ee 
where we set 
\be\la{rrr2} 
\si(\x):= \sum_j
\si_j (\x-\x_j), 
\qquad \nu(\x):=e|\psi(\x)|^2
\ee 
similarly to (\re{rrr}). Let us note that the charge densities 
$\si$ and $\rho$  are finite Borelian measures on $T^3$
for $\psi\in H^1$ since $\psi\in L^6$ by the Sobolev embedding theorem.

For sufficiently smooth (smeared) ion densities $\si(\x)$ 
 the energy in the periodic cell is defined as in \ci{K-2014}:
\be\la{HamsT} 
E(\psi,\ov\x):=
\fr{\h^2}{2\cm}\langle \na\psi(\x), \na\psi(\x)\rangle
+\fr12 \langle \phi , \rho  \rangle,\qquad\qquad \phi:=  Q\rho 
\ee 
where 
$\langle\cdot ,\cdot \rangle$ denotes the Hermitian scalar product in $L^2$, and 
$Q\rho:=(-\De)^{-1}\rho$ is well-defined by (\re{neu1}). 
Namely, consider the dual lattice 
\be\la{Ga3d} 
\Ga^*=\{\bk(\n)=\bb_1n_1+\bb_2n_2+\bb_3n_3: \n=(n_1,n_2,n_3)\in\Z^3  \}, 
\ee 
where 
$\bb_k\bba_{k'}=2\pi\de_{kk'}$. 
Every finite measure $\rho$ on $T^3$ admits the Fourier representation
\be\la{Fou} 
\rho(\x)=\fr1{\sqrt{|T^3|}}\sum_{\bk\in\Ga^*}\hat\rho(\bk) e^{-i\bk\x},\qquad 
\hat\rho(\bk)=\fr1{\sqrt{|T^3|}}\int_{T^3} e^{i\bk\x}\rho(\x)d\x.
\ee 
where the Fourier coefficients $\hat\rho(\bk)$  are bounded.
Respectively, we define the Coulomb potential
\be\la{Fou2} 
\phi(\x)= 
Q\rho(\x) 
:= \fr1{\sqrt{|T^3|}}\sum_{\bk\in\Ga^*\setminus 0}\fr{\hat\rho(\bk)}{\bk^2} 
e^{-i\bk\x}. 
\ee 
This function 
$\phi\in L^2$
and satisfies 
the Poisson equation $-\De\phi=\rho$, 
since $\hat\rho(0)=0$ 
due to the neutrality condition (\re{neu1}).
Finally,
\be\la{Fou3} 
\int_{T^3}\phi(\x)d\x=0. 
\ee 
For the smeared ions 
the energy  (\re{HamsT}) can be rewritten as 
\be\la{HamsTf} 
E(\psi,\ov\x)=
\fr{\h^2}{2\cm}\langle \na\psi, \na\psi\rangle
+ 
\fr12\langle  Q\si , \si  \rangle+\langle  Q\si , \nu  \rangle+\fr12 \langle  Q\nu , \nu  \rangle.
\ee 
Let us show that 
for the point ions
the Coulomb  selfaction energy $\langle  Q\si, \si\rangle=\sum_{j,k=1}^N \langle Q\si_j, \si_k\rangle$ is infinite.
Namely, according to (\re{Fou2}), the Coulomb potential of the ions reads
\be\la{Fou2n}  
\phi_{\rm ions}(\x):=Q\si(\x) 
= \fr1{\sqrt{|T^3|}}\sum_{\bk\in\Ga^*\setminus 0}\fr{\hat\si(\bk)}{\bk^2},\qquad\qquad \hat\si(\bk)=
\fr{|e|}{\sqrt{|T^3|}}\sum_j Z_j e^{i\bk\x_j}. 
\ee 
Hence, for the point ions
\be\la{Fou2h}  
\phi_{\rm ions}(\x)=\sum_j \phi_j(\x),\qquad  \phi_j(\x):=Q\si_j(\x-\x_j) =|e| Z_j G(\x-\x_j),
 \qquad
G(\x)=\sum_{\bk\in\Ga^*\setminus 0}\fr{e^{-i\bk\x}} {\bk^2}
\ee 
where $G(\x)$ is the Green function introduced in  \ci{CBL1998}. Obviously, $\ds\int_{T^3}G(\x)d\x=0$, and 
$-\De G(\x)=\de(\x)$. Therefore, by the elliptic regularity,  
\be\la{G}
G\in C^\infty(T^3\setminus 0),\qquad \qquad D(\x):=G(\x)-\fr 1{4\pi |\x|}\in C^\infty(|\x|<\ve)
\ee
for sufficiently small $\ve>0$.
As the result, the selfaction terms
$
\langle Q\si_j(\x-\x_j), \si_j(\x-\x_j)\rangle=|e|^2 Z_j^2 G(0)
$
 are infinite, while 
$\langle Q\si_j(\x-\x_j), \si_k(\x-\x_k)\rangle=|e|^2 Z_jZ_k G(\x_j-\x_k)$ are 
finite for $j\ne k$.

\br\la{rD0}
Let us note that $G(\x)$ is symmetric
with respect to the reflection $\x\mapsto -\x$ of the torus $T^3$.
Therefore, the difference $D(\x)$ is symmetric in the ball $|\x|<\ve$ with respect to this reflection, and hence
\be\la{G0}
\na D(0)=0.
\ee
\er
From now on
we consider the point ions (\re{rrr}), and we
will
renormalize the energy (\re{HamsTf}) subtracting the infinite selfaction terms: 
\be\la{HamsTr} 
E_r(\psi,\ov\x)=
\fr{\h^2}{2\cm}\langle \na\psi, \na\psi\rangle
+ 
\fr12\sum_{j\ne k}
\langle Q\si_j(\x-\x_j), \si_k(\x-\x_k)\rangle
+\langle  Q\si , \nu  \rangle+\fr12 \langle  Q\nu , \nu  \rangle.
\ee 
Let us note that $\nu\in L^2$ 
for $\psi\in H^1$
by the Sobolev embedding theorem, and 
$Q\si\in L^2$.
Hence, 
the renormalized energy  is finite 
for $\psi\in H^1$.
Next problem is to check that the  renormalized energy is bounded from below.
Let us denote 
\be\la{cX}
\cX:=\{\ov\x\in (T^3)^N: \x_j\ne \x_k ~~\mbox{\rm for}~~j\ne k\}, \qquad d(\ov\x):=\min_{j\ne k} \dist (\x_j,\x_k).
\ee

\bd $\cM:=M \times \cX$,
where $M$ denotes the manifold (cf. (\re{neuZ}))
\be\la{MZ} 
M=\{\psi\in H^1:~ \int_{T^3} |\psi(\x)|^2d\x 
=Z \}
\ee 
endowed with the topology of $H^1\times \cX$.
\ed

\bl\la{lf}
Let condition (\re{Zp}) hold. Then
the functional $E_r$ is continuous on $\cM$, and the bound holds 
\be\la{HamsTr2} 
E_r(\psi,\ov\x)\ge \ve\Vert\psi\Vert_{H^1}^2 
+\fr{q}{d(\ov\x)}
+\fr12 \langle Q\nu,\nu\rangle-C,\qquad (\psi,\ov\x)\in \cM,
\ee
where $q,\ve>0$.
\el
\Pr
First, $\nu:=e|\psi(\x)|^2\in L^2$ since $\Vert\nu\Vert_{L^2}=e^2\Vert\psi\Vert_{L^4}^2 
\le C_1\Vert\psi\Vert_{H^1}^2$
by the Sobolev embedding theorem \ci{Adams, Sob}. Further, $Q\si\in L^2$ since $\si$
is the finite Borelian measure on $T^3$ by (\re{rrr2}). 
Hence, for any $\de>0$ 
\be\la{f}
|\langle Q\si(\x),\nu(\x)\rangle| \le C\Vert\psi\Vert_{L^4}^2\le  \de\Vert\psi\Vert_{L^6}^2+C(\de)  \Vert\psi\Vert_{L^2}^2  \le 
C_2\de \Vert\psi\Vert_{H^1}^2+C(\de) Z.
\ee
Here the second inequality follows by  the Young inequality
from
$\Vert\psi\Vert_{L^4}\le \Vert\psi\Vert_{L^6}^{3/4}\Vert\psi\Vert_{L^2}^{1/4}$
which holds
by the Riesz convexity theorem. This theorem  follows by the H\"older
inequality, and in our case the Cauchy-Schwarz is sufficient:
$$
\int |\psi(\x)|^3  |\psi(\x)|   d\x\le [\int |\psi(\x)|^6d\x]^{1/2} 
[\int |\psi(\x)|^2d\x]^{1/2}.
$$ 
Therefore, the functional $(\psi,\ov\x)\mapsto\langle Q\si,\nu\rangle$ 
is continuous on $\cM$ in the topology of $H^1\times \cX$.

On the other hand, for $\psi\in M$ we have
$\Vert\psi\Vert_{H^1}^2=
\ds\int_{T^3} |\na\psi(\x)|^2 d\x +Z$. Hence, the bound (\re{HamsTr2}) follows if we take $C_2\de<\fr{\h^2}{2\cm}$.\bo

\setcounter{subsection}{0} 
\setcounter{theorem}{0} 
\setcounter{equation}{0}

\section{Compactness of minimizing sequence}

The energy is finite and bounded from below on the manifold $\cM$ by Lemma \re{lf}.
Hence, there exists a minimizing sequence $(\psi_n, \ov\x_n)\in\mathcal M$ such that 
\be\la{min} 
E_r(\psi_n,\ov\x_n)\to E^0_r:=\inf_\mathcal M~ 
E(\psi,\ov\x), \qquad n\to\infty. 
\ee 
\br
For sufficiently smooth charge densitites $\si_j$ the energy (\re{HamsTf}) is also finite, and its difference 
with
(\re{HamsTr}) equals $\fr12 \sum \langle Q\si_j(\x-\x_j), \si_j(\x-\x_j)\rangle=\fr12 \sum \langle Q\si_j, \si_j\rangle$.
This difference does not depend on $\psi$ and $\ov\x$. Hence, 
the corresponding minimizers coincide.
\er

Our main result is the following: 
\bt\la{t3} 
i) There exists  $(\psi^0,\ov\x^0)\in \mathcal M$ with
\be\la{U0min} 
E_r(\psi^0,\ov\x^0)= E^0_r. 
\ee 
ii) Moreover,  
$\psi^0$ 
 satisfies equations \eqref{LPS3}--\eqref{LPS3g} with a real
potential $\phi^0\in L^2$ 
 and  $\om^0\in\R$. 

\et 
To prove i),
let us denote 
\be\la{ron}
\rho_n(\x):=\si_n(\x)+e|\psi_n(\x)|^2,\qquad \si_n(\x):=\sum_{j}\mu^{\rm per}_{j}(\x-\x_{jn}),
\qquad
\nu_n(\x):=e|\psi_n(\x)|^2.
\ee
The sequence $\psi_n$  is bounded in $H^1$ 
by (\re{min}) and (\re{HamsTr2}), and hence the corresponding sequence $\nu_n$ is bounded
in $L^2$ by the Sobolev embedding theorem \ci{Adams,Sob}. 
Respectively, the corresponding sequences $Q\si_n$ and $\phi_n:=Q\rho_n$ 
are  bounded in $L^2$.

Hence, the sequence $\psi_n$ is precompact in~$L^p$ 
for any $p\in[1,6)$ by the Sobolev embedding theorem. 
As the result,
there exist a  subsequence $n'\to\infty$ for which  
\be\la{subs} 
\psi_{n'}\toLp\psi^0, \quad 
\nu_{n'}(\x) \toLd\nu^0, \quad
\phi_{n'}\tow\phi^0, \quad
\ov\x_{n'}\to \ov\x^0 \in\cX, \qquad \qquad n'\to\infty
\ee 
with any $p\in[1,6)$.
Respectively, the convergences
\be\la{subs1} 
\si_{n'}\to \si^0,  \qquad \rho_{n'}\to \rho^0, \qquad \qquad n'\to\infty.
\ee 
hold in the sense of distributions,
where $\si^0(\x)$ and $\rho^0(\x)$ are defined by (\re{rrr}) and (\re{LPS4}). 
Therefore, 
\be\la{3subs2} 
Q\si_{n'}\tow Q\si^0, \qquad \qquad n'\to\infty.
\ee 
Hence, the neutrality condition (\re{neu10}) holds, 
$(\psi^0,\ov\x^0)\in \mathcal M$, $\phi^0\in L^2$, and for these limit functions we have
\be\la{phi0}
-\De\phi^0=\rho^0, \qquad \ds\int_{{T^3}}\phi^0(\x)d\x=0.
\ee 
To prove identity (\re{U0min}), we write the energy (\re{HamsTr}) as the sum 
$E_r=E_1+E_2+E_3+E_4$, where
$$
\ba{rclrcl}
E_1(\psi,\ov\x)&=&\fr{\h^2}{2\cm}
\langle\na\psi(\x),\na\psi(\x)\rangle,&\qquad
E_2(\psi,\ov\x)&=&\fr12\sum_{j\ne k} \langle Q\si(\x-\x_j),  \si(\x-\x_k) \rangle,
\nonumber\\
\nonumber\\
\nonumber
E_3(\psi,\ov\x)&=&
\langle Q\si(\x),\nu(\x)\rangle, &
E_4(\psi,\ov\x)&=&
\fr12\langle Q\nu(\x),\nu(\x)\rangle. ~~~~
\ea 
$$
Finally,
the convergences (\re{subs}) and (\re{3subs2}) imply that
$$
E_1(\psi^0,\ov\x^0)\le \liminf_{n'\to\infty} E_1(\psi_{n'},\ov\x_{n'}), \qquad\qquad  E_l(\psi^0,\ov\x^0)= \lim_{n'\to\infty} E_l(\psi_{n'},\ov\x_{n'}),
 \quad l=2,3,4.
$$
These limits, together with (\re{min}),  give that 
$
E_r(\psi^0,\ov\x^0) 
\le E^0_r. 
$
Now (\re{U0min}) follows from the definition of $E^0_r$, since 
$(\psi^0,\ov\x^0)\in\mathcal M$. Thus Theorem \re{t3} i) is proved.
\medskip

We will prove Theorem \re{t3} ii) in next sections.

\setcounter{subsection}{0} 
\setcounter{theorem}{0} 
\setcounter{equation}{0}

\section{Variation of the energy} 

Theorem \re{t3} ii) follows from next proposition.

\bp\la{tgs23} The limit functions 
(\re{subs})
satisfy equations {\rm \eqref{LPS3}--\eqref{LPS3g}} with  $\om^0\in\R$.

\ep

The Poisson equation (\re{LPS4}) is proved in (\re{phi0}). The Lorentz equation
 \eqref{LPS3g} follows by differentiation of the energy (\re{HamsTr}) in $\x_j$. 
 Namely, the derivative at the minimal point  $(\psi^0,\ov\x^0)$ should vanish: taking into account (\re{Fou2h}), we obtain
 $$
 \ba{rcl}
 0&=& \na_{\x_j}E_r(\psi^0,\ov\x^0)=\sum_{k\ne j}\langle Q\na\si_j(\x-\x_j^0),\si_k(\x-\x_k^0) \rangle
 +\langle Q\na\si_j(\x-\x_j^0),\nu^0 \rangle
 \nonumber\\
 \nonumber\\
 \nonumber
 &=&
 \langle \na\si_j(\x-\x_j^0),\phi^0(\x)-\phi_j^0(\x) \rangle=-\langle \si_j(\x-\x_j^0),\na[\phi^0(\x)-\phi_j^0(\x)] \rangle,
 \ea
 $$
 where $\phi_j^0(\x):=Q\si_j(\x-\x_j^0)$ similarly to (\re{Fou2h}).
 Finally, 
 the last expression coincides with the right hand side of \eqref{LPS3g} by 
 its definition (\re{phinj0})
together with (\re{G0}).

It remains to prove the Schr\"odinger equation \eqref{LPS3}. 
Let us denote $\E_r(\psi):=E_r(\psi,\ov\x^0)$.
We derive \eqref{LPS3} in next sections, 
equating the variation of $\E_r(\cdot)|_{M}$ to zero at $\psi=\psi^0$. 
In this section we calculate the corresponding G\^ateaux variational derivative.

We should work directly on~$M$ introducing an atlas in a~neighborhood 
of $\psi^0$ in $M$.
We define the atlas 
as the stereographic projection from the tangent plane 
$TM(\psi^0)=(\psi^0)^\bot:=\{\psi\in H^1: 
\langle\psi,\psi^0\rangle=0\}$ 
to the sphere (\re{MZ}): 
\be\la{3atlas} 
\psi_\tau= \fr{\psi^0+\tau~~~}{\Vert\psi^0+\tau\Vert_{L^2}} 
\sqrt{Z}, \qquad  \tau\in (\psi^0)^\bot. 
\ee 
Obviously, 
\be\la{3tau} 
\fr d{d\ve}\Big|_{\ve=0} \psi_{\ve\tau}=\tau,\qquad \tau\in (\psi^0)^\bot, 
\ee 
where the derivative exists in $H^1$. We define
the 'G\^ateaux derivative' of $\E_r(\cdot)|_{M}$  as 
\be\la{3Gder} 
D_\tau \E_r(\psi^0):=\lim_{\ve\to 0}\fr{\E_r(\psi_{\ve\tau})-\E_r(\psi^0)}{\ve},
\ee 
if this limit exists. We should restrict the set of allowed tangent vectors~$\tau$. 
 
\bd 
$\ccT^0$ is the space of test functions 
$\tau\in(\psi^0)^\bot\cap C^\infty(T^3)$. 
\ed 
Obviously, $\ccT^0$ is dense in $(\psi^0)^\bot$ in the norm of  $H^1$. 
 
%%%%%%%%%%%%%%%%%%%%%%%%%%%%%%%%%%%%%%%%%%%%
 
 %%%%%%%%%%% XI
 
 %%%%%%%%%%%%%%%%%%%%%%%%%%%%%%%%%%%%%%%%%
 
\bl\la{3lvar} 
Let 
$\tau\in\ccT^0$. Then the  derivative \eqref{3Gder} exists, and
\be\la{3Gder2} 
D_\tau \E_r(\psi^0)=\int_{T^3}  \Big[\fr{\h^2}{2\cm} 
(\na\tau\ov{\na\psi^0}+\na\psi^0\ov{\na \tau}) 
+e {Q\rho^0}(\tau\ov{\psi^0}+\psi^0\ov \tau )\Big]d\x. 
\ee 
\el 
\Pr 
Let us denote $\nu_{\ve\tau}(\x):=e|\psi_{\ve\tau}(\x)|^2$.

\bl\la{3lL2}
 For $\tau\in\ccT^0$ we have $\nu_{\ve\tau}\in L^2$, and
 \be\la{3Gder3} 
D_\tau \nu:= 
\lim_{\ve\to 0}\fr{\nu_{\ve\tau}-\nu_0}{\ve} 
 =e(\tau\ov{\psi^0}+\psi^0\ov \tau ), 
\ee 
where the limit converges in  $L^2$.

\el
\Pr In the polar coordinates
\be\la{3al} 
\psi_{\ve\tau}=(\psi^0+\ve\tau)\cos\al,\qquad 
\al=\al(\ve)=\arctan\fr{\ve\Vert\tau\Vert_{L^2}}{\Vert\psi^0\Vert_{L^2}}. 
\ee 
Hence,
\beqn\la{3rod} 
\nu_{\ve\tau}&=&e\cos^2\al|\psi^0+\ve\tau|^2 
\nonumber\\
\nonumber\smallskip\\
&=& 
\nu^0+e\ve\cos^2\al(\tau\ov{\psi^0}+\psi^0\ov\tau)+ 
e[\ve^2|\tau|^2\cos^2\al-|\psi^0|^2\sin^2\al]. 
\eeqn 
It remains to 
estimate 
 the last term 
of (\re{3rod}),
\be\la{3lt}
R_\ve:=\Lam[\ve^2|\tau|^2\cos^2\al-|\psi^0|^2\sin^2\al].
\ee
Here $|\psi^0|^2\in L^2$ 
since $\psi^0\in H^1\subset L^6$. Finally,  $|\tau|^2\in L^2$ and $\sin^2\al\sim \ve^2$.
Hence, the convergence (\re{3Gder3}) holds in $L^2$.\bo
\medskip

Now (\re{3Gder2}) follows by differentiation 
in $\ve$ of (\re{HamsTr}) with $\psi=\psi_{\ve\tau}$, $\si=\si^0$ and $\nu=\nu_{\ve\tau}$.
\bo

\setcounter{subsection}{0} 
\setcounter{theorem}{0} 
\setcounter{equation}{0}

\section{The Schr\"odinger equation}

Since $\psi^0$ is a minimal point, the G\^ateaux derivative (\re{3Gder2}) 
vanishes: 
\be\la{3GaD} 
\int_{T_2}  \Big[\fr{\h^2}{2\cm} 
(\na\tau\ov{\na\psi^0}+\na\psi^0\ov{\na \tau}) 
+
e {Q\rho^0}(\tau\ov{\psi^0}+\psi^0\ov \tau )
\Big]d\x=0. 
\ee 
Substituting $i\tau$ instead of $\tau$ in this identity 
and subtracting, we obtain 
\be\la{3GaD2} 
-\fr{\h^2}{2\cm}\langle\De\psi^0, \tau\rangle 
+ e \langle 
Q\rho^0,\ov{\psi^0} \tau \rangle=0. 
\ee 
Finally,
\be\la{3sp} 
\langle 
Q\rho^0,\ov{\psi^0} \tau \rangle= 
\langle\phi^0\psi^0, \tau \rangle
\ee 
since
$\rho^0=-\De\phi^0$. 
Hence, we can rewrite (\re{3GaD2}) as the variational identity 
\be\la{3GaD22} 
\langle -\fr{\h^2}{2\cm}\De\psi^0 
+ e 
\phi^0\psi^0,\tau \rangle=0,    \qquad \tau\in\ccT^0. 
\ee 
Now we can prove the Schr\"odinger equation (\re{LPS3}).
\bl\la{lse}
$\psi^0$ is the eigenfunction of the 
Schr\"odinger operator $H=-\fr{\h^2}{2\cm}\De+e\phi^0$:
\be\la{3Hpsi} 
H\psi^0=\lam\psi^0,
\ee 
where 
$\lam\in\R$. 
\el
\Pr
First, $H\psi^0$ is a well-defined  distribution since  $\phi^0,\psi^0\in L^2$ by (\re{subs}), and hence   $\phi^0\psi^0\in L^1$. Second,
 $\psi^0\ne 0$ since $\psi^0\in M$ and $Z>0$. Hence, there exists a~test function 
$\theta\in C^\infty(T^3)\setminus\ccT^0$, i.e., 
\be\la{3test} 
\langle\psi^0,\theta\rangle\ne 0. 
\ee 
Then 
\be\la{3test2} 
\langle(H-\lam)\psi^0,\theta\rangle= 0. 
\ee 
for an appropriate $\lam\in\C$. However, 
$(H-\lam)\psi^0$ also  annihilates $\ccT^0$ by (\re{3GaD22}), 
hence it  annihilates the whole space 
$C^\infty(T^3)$. This implies (\re{3Hpsi}) in the sense of 
distributions with a $\lam\in\C$. Finally, (\re{3Hpsi}) gives
\be\la{3Hpsi2} 
\langle H\psi^0,\psi^0\rangle=\lam\langle\psi^0,\psi^0\rangle,
\ee 
where the left hand side is well defined since $\psi^0\in H^1$ and $\psi^0\in L^4$,
while  $\phi^0\in L^2$. Therefore, $\lam\in\R$ since 
the potential is  real.\bo
 \medskip

 This lemma implies  equation (\re{LPS3})  with $\hbar\om^0=\lam$, and hence 
Theorem \re{t3} ii) is proved.

\end{document}